\documentclass{ws-procs9x6}

\begin{document}

\title{FULLY MICROSCOPIC CALCULATIONS FOR CLOSED-SHELL NUCLEI 
 WITH REALISTIC NUCLEON-NUCLEON POTENTIALS}

\author{L. CORAGGIO, A. COVELLO, A. GARGANO, and N. ITACO}

\address{Dipartimento di Scienze Fisiche, Universit\`a
di Napoli Federico II, \\ and Istituto Nazionale di Fisica Nucleare, \\
Complesso Universitario di Monte  S. Angelo, Via Cintia - I-80126 Napoli,
Italy}

\begin{abstract}
The ground-state energy of the doubly magic nuclei $^4$He and $^{16}$O
has been calculated within the framework of the Goldstone expansion
starting from modern nucleon-nucleon potentials. 
A low-momentum potential $V_{\rm low-k}$ has been derived from the
bare potential by integrating out its high-momentum components
beyond a cutoff $\Lambda$. 
We have employed a simple criterion to relate this cutoff momentum to
a boundary condition for the two-nucleon model space spanned by a
harmonic-oscillator basis.  
Convergence of the results has been obtained with a limited
number of oscillator quanta. 
\end{abstract}

\bodymatter

\section{Introduction}
As is well known, the strong repulsive components in the high-momentum
regime of a realistic nucleon-nucleon ($NN$) potential $V_{NN}$ need
to be renormalized in order to perform perturbative nuclear structure
calculations. 
In Refs. \cite{bogner01,bogner02} a new method to renormalize the $NN$
interaction has been proposed, which consists in deriving an effective
low-momentum potential $V_{\rm low-k}$ that satisfies a decoupling
condition between the low- and high-momentum spaces. 
This $V_{\rm low-k}$ preserves exactly the on-shell properties of the
original $V_{NN}$ up to a cutoff momentum $\Lambda$, and is a smooth
potential which can be used directly in nuclear structure calculations. 

In the past few years, we have employed this approach to calculate
the ground-state (g.s.) properties of doubly closed-shell nuclei 
within the framework of the Goldstone expansion
\cite{coraggio03,coraggio05}, using a fixed value of the cutoff
momentum. 

Recently, we have investigated how the cutoff momentum $\Lambda$ is
related to the dimension of the configuration space in the coordinate
representation \cite{coraggio06}, where our calculations are
performed. 
We have shown how the choice of a cutoff momentum corresponds to fix a
boundary for the two-nucleon model space.

In the present work, we calculate the g.s. energy of $^{4}$He and
$^{16}$O in the framework of the Goldstone expansion with different
$NN$ potentials. 
To verify the validity of our approach, we compare the $^{4}$He
results with those obtained using the Faddeev-Yakubovsky (FY) method.

The paper is organized as follows. 
In Sec. 2 we give a brief description of our calculations.
Sec. 3 is devoted to the presentation and discussion of our
results for $^{4}$He and $^{16}$O. 
A summary of our study is given in Sec. 4.

\section{Outline of calculations}
As mentioned in the Introduction, the short-range repulsion of the
$NN$ potential is renormalized integrating out its high-momentum
components through the so-called $V_{\rm low-k}$ approach (see
Refs. \cite{bogner01,bogner02}). 
The $V_{\rm low-k}$ preserves the physics of the two-nucleon system up
to the cutoff momentum $\Lambda$. 
While this low-momentum potential is defined in the momentum space, we
perform our calculations for finite nuclei in the coordinate space
employing a truncated HO basis. 
This makes it desirable to map the cutoff momentum $\Lambda$, which
decouples the momentum space into a low- and high-momentum regime,
onto a boundary for the HO space \cite{coraggio06}.

If we consider the two-nucleon relative motion in a HO well in the
momentum representation, then, for a given maximum relative momentum
$\Lambda$, the corresponding maximum value of the energy is 

\begin{equation}
E_{\rm max} = \frac{ \hbar^2 \Lambda^2}{M}~~,
\label{one}
\end{equation}

\noindent
where $M$ is the nucleon mass.

We rewrite this relation in the relative coordinate system in terms of
the maximum number $N_{\rm max}$ of HO quanta:

\begin{equation}
\left( N_{\rm max} + \frac{3}{2} \right) \hbar \omega = \frac{ 
\hbar^2 \Lambda^2}{M}~~,
\label{two}
\end{equation}

\noindent
for a given HO parameter $\hbar \omega$.
The above equation provides a simple criterion to map out the
two-nucleon HO model space.
Let us write the two-nucleon states as the product of HO wave functions

\begin{equation}
|a~b \rangle = | n_a l_a j_a,~n_b l_b j_b \rangle~~.
\label{three}
\end{equation}

\noindent
We define our HO model space as spanned by those two-nucleon states that
satisfy the constraint

\begin{equation}
2n_a+l_a+2n_b+l_b \leq N_{\rm max}~~.
\label{four}
\end{equation}

In this paper, making use of the above approach, we have calculated
the g.s. energies of $^{4}$He and $^{16}$O within the framework of the
Goldstone expansion \cite{goldstone57}. 
We start from the intrinsic hamiltonian 

\begin{equation}
H= \left( 1 - \frac{1}{A} \right) \sum_{i=1}^{A} \frac{p_i^2}{2M} + \sum_{i<j}
\left( V_{ij} - \frac{ {\rm {\bf p}}_i \cdot {\rm {\bf p}}_j }{MA} \right)~~, 
\end{equation}

\noindent
where $V_{ij}$ stands for the renormalized $V_{NN}$ potential plus the 
Coulomb force, and construct the Hartree-Fock (HF) basis expanding the
HF single particle (SP) states in terms of HO wave functions.
The following step is to sum up the Goldstone expansion including 
contributions up to fourth-order in the two-body interaction.
Using Pad\'e approximants \cite{baker70,ayoub79} one may obtain a value
to which the perturbation series should converge.
In this work, we report results obtained using the Pad\'e approximant
$[2|2]$, whose explicit expression is

\begin{equation}
[ 2|2 ]=
\frac{E_0(1+\gamma_1+\gamma_2)+E_1(1+\gamma_2)+E_2}{1+\gamma_1+
\gamma_2}~~,
\end{equation}

\noindent
where 

\[
\gamma_1 = \frac{E_2E_4-E_3^2}{E_1E_3-E_2^2} ~~,~~~~~
\gamma_2 = -\frac{E_3+E_1 \gamma_1}{E_2} \nonumber ~~,
\]

\noindent
$E_i$ being the $i$th order energy contribution in the Goldstone
expansion.

Our calculations are made in a truncated model space, whose size 
is related to the values of the cutoff momentum $\Lambda$ and the $\hbar 
\omega$ parameter.
The calculations are performed increasing the $N_{\rm max}$ value
(and consequently $\Lambda$) and varying $\hbar \omega$ until the
dependence on $N_{\rm max}$ ($\Lambda$) is minimized.

\section{Results}

\begin{figure}
\psfig{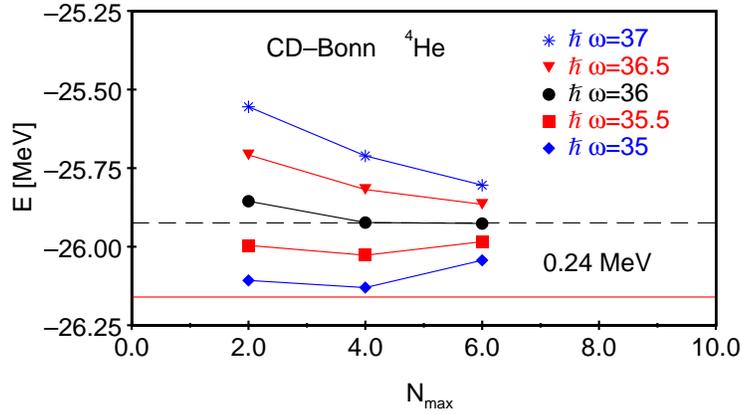}
\caption{Ground state energy of $^4$He calculated with the CD-Bonn
  potential as function of $N_{\rm max}$, for different values of 
$\hbar \omega$. 
The straight line represents the Faddeev-Yakubovsky result, while the
  dashed one our converged result.
The difference in energy between the latter and the Faddeev-Yakubovsky
  result is also reported.}
\label{4hecdbonn}
\end{figure}

\begin{figure}
\psfig{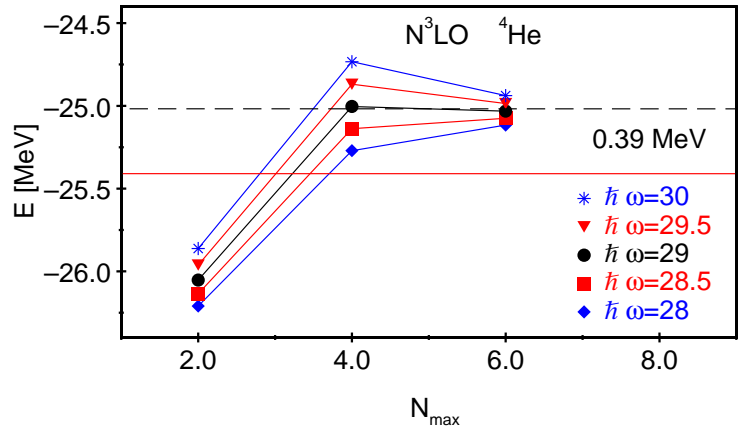}
\caption{Same as Fig. \ref{4hecdbonn}, but for the N$^3$LO potential.}
\label{4heN3LO}
\end{figure}

\begin{figure}
\psfig{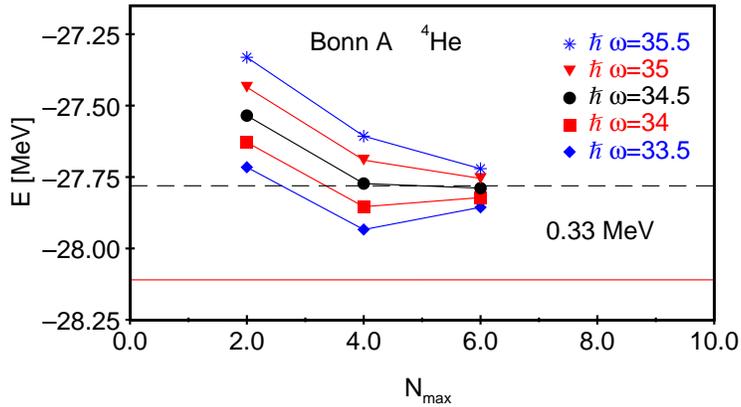}
\caption{Same as Fig. \ref{4hecdbonn}, but for the Bonn A potential.}
\label{4hebonna}
\end{figure}

We have calculated the binding energy of $^4$He using different
$V_{NN}$'s, and compared our results with those obtained by means of
the FY method. 
This comparison is made in order to test the reliability of our
approach.
In Figs. \ref{4hecdbonn}, \ref{4heN3LO}, and
\ref{4hebonna} the calculated $^4$He g.s. energies obtained from the
CD-Bonn \cite{cdbonn01}, N$^3$LO \cite{entem03}, and
Bonn A \cite{mach87} $NN$ potentials are reported, for different
values of $\hbar \omega$, as a function of the maximum number $N_{\rm
  max}$ of HO quanta.
The FY result \cite{gloeckle93,nogga05} is also shown.

For the sake of clarity, in Table \ref{tablecdbonn} we report the
numerical values obtained with the CD-Bonn potential. 
From the inspection of Table \ref{tablecdbonn} it can be seen that
the g.s. energy does not change increasing $N_{\rm max}$ from 4 to 6
for $\hbar \omega=36$ MeV.
On these grounds, we choose as our final result that corresponding
to the above $\hbar \omega$ value, i.e. -25.92 MeV.
Moreover, we find it worthwhile to introduce a theoretical error due to
the uncertainty when choosing $\hbar \omega_{\rm conv}$, which
corresponds to the one with the faster convergence with $N_{\rm max}$. 
We estimate this error as the largest difference in energy between the
final result and those corresponding to the two $\hbar \omega$
values adjacent to $\hbar \omega_{\rm conv}$. 
For the CD-Bonn potential, we see that this difference is 0.05 MeV for
the largest $N_{\rm max}$. 

\begin{table}
\tbl{Ground state energy of $^4$He (in MeV) calculated with the
  CD-Bonn potential for different values of $\hbar
  \omega$ and $N_{\rm max}$}
{\begin{tabular}{@{}cccc@{}}\toprule
$\hbar \omega$ & $N_{\rm max}=2$ & $N_{\rm max}=4$ & $N_{\rm max}=6$ \\
(in MeV)       & ~ & ~& ~ \\\colrule
35.0 & -26.11 & -26.13 & -26.04 \\
35.5 & -25.10 & -26.03 & -25.97 \\
36.0 & -25.86 & -25.92 & -25.92 \\
36.5 & -25.71 & -25.83 & -25.87 \\
37.0 & -25.56 & -25.71 & -25.80 \\\botrule
\end{tabular}}
\label{tablecdbonn}
\end{table}

Similarly, the results for the $^4$He g.s. energy with the N$^3$LO and
the Bonn A potentials are ($-25.02 \pm 0.05$) and ($-27.78 \pm 0.03$)
MeV, respectively. 
These values are in good agreement with the FY results, the largest 
discrepancy being 0.39 MeV for N$^3$LO potential.

\begin{figure}
\psfig{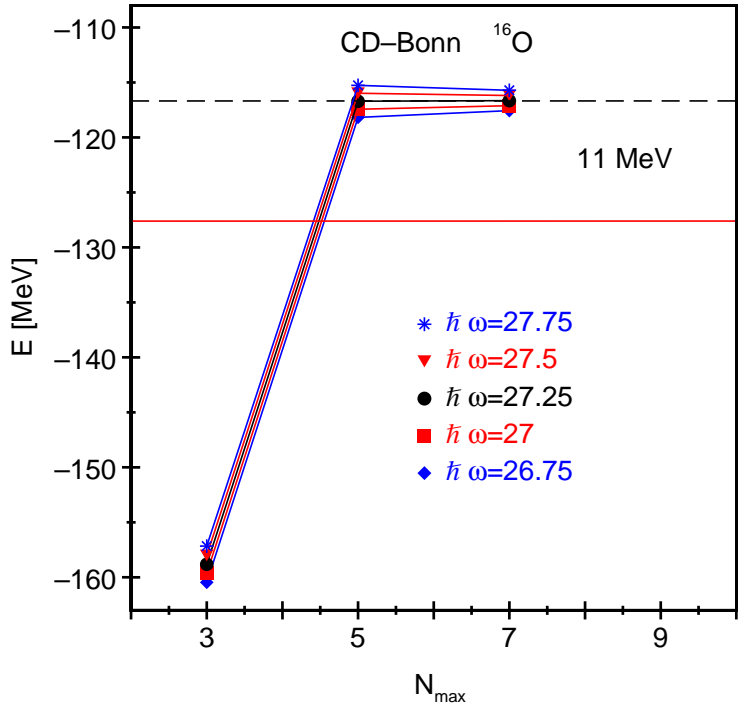}
\caption{Ground state energy of $^{16}$O calculated with the CD-Bonn 
potential as function of $N_{\rm max}$, for different values of $\hbar 
\omega$. 
The straight line represents the experimental value \cite{audi03}, while the
  dashed one our converged result.
The difference in energy between the latter and the experimental value is
  also reported.}
\label{16ocdbonn}
\end{figure}

\begin{figure}
\psfig{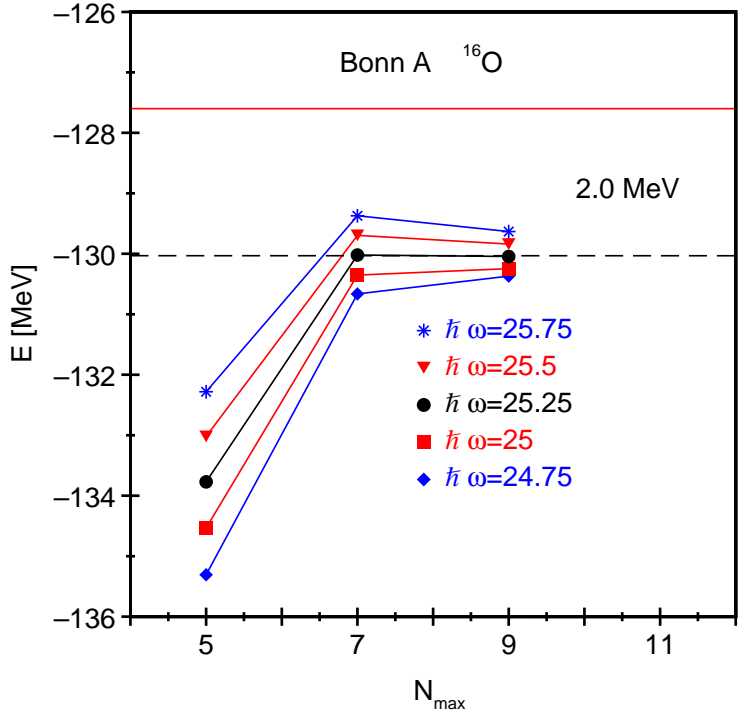}
\caption{Same as Fig. \ref{16ocdbonn}, but for the Bonn A potential.}
\label{16obonna}
\end{figure}

We have also calculated the g.s. energy of $^{16}$O starting from both
the CD-Bonn and the Bonn A potential, as reported in Figs.
\ref{16ocdbonn} and \ref{16obonna}, respectively. 
With the CD-Bonn potential, the converged value, obtained for $\hbar
\omega=$ 27.25 MeV, is equal to ($-117 \pm 1$) MeV, the discrepancy
with the experimental value \cite{audi03} being 11 MeV. 
This value is slightly different ($\approx 1$ MeV) from the one
reported in our previous paper \cite{coraggio06}, because in the present
work we have decreased by a factor 2 the spacings between the $\hbar
\omega$ values. 
It is worth to point out that this result is consistent with those 
obtained by Fujii {\it et al.} using the unitary model-operator
approach \cite{fujii05}, and by Vary {\it et al.} in the no-core shell
model framework \cite{vary05}.

A better agreement with experiment is obtained using the weaker
tensor force $NN$ potential Bonn A, our $^{16}$O g.s. energy being
($-130.0 \pm 0.5$) MeV.

\section{Summary}
In this work, we have calculated the g.s. energy of the doubly
closed-shell nuclei $^{4}$He and $^{16}$O in the framework of the
Goldstone expansion, starting from different realistic $NN$
potentials. 
In order to renormalize their short-range repulsion, the high-momentum
components of these potentials have been integrated out through the
so-called $V_{\rm low-k}$ approach. 
We have employed a criterion to map out the model space of the
two-nucleon states in the HO basis according to the value of the
cutoff momentum $\Lambda$ \cite{coraggio06}. 

To show the validity of this procedure, we have calculated the
g.s. energy of $^4$He, with the CD-Bonn, N$^3$LO, and Bonn A
potentials, comparing the results with the FY ones.
We have found that the energy differences do not exceed 0.39 MeV. 
The limited size of the discrepancies evidences that our approach 
provides a reliable way to renormalize the $NN$ potentials preserving
not only the two-body  but also the many-body physics.

On the above grounds, we have performed similar calculations for
$^{16}$O with the CD-Bonn and Bonn A $NN$ potentials, and obtained
converged results using  model spaces not exceeding $N_{\rm max}=9$. 
The rapid convergence of the results with the size of the HO model
space makes it very interesting to study heavier systems employing our
approach \cite{coraggio07b}.

\section{Acknowledgments}
This work was supported in part by the Italian Ministero
dell'Istruzione, dell'Universit\`a e della Ricerca (MIUR).

\end{document}